\documentclass[a4paper,11pt]{article}
\pdfoutput=1 

\usepackage{jcappub} 

\usepackage[T1]{fontenc} 
\usepackage{graphicx}
\newcommand{\hmpc}{{\, h^{-1}\, {\rm Mpc}}}

\def\aj{AJ}
\def\apj{ApJ}
\def\apjs{ApJS}
\def\jcap{JCAP}
\def\mnras{MNRAS}
\def\aap{A\&A}

\def\nat{Nature}      
\def\apjs{ApJS}
\def\apjl{ApJ Letters}
\def\physrep{Physics Reports}

\title{\boldmath Renyi entropy as a measure of cosmic homogeneity}

\author[a]{Biswajit Pandey}

\affiliation[a]{Department of Physics, Visva-Bharati University,
  Santiniketan, 731235, India}
\emailAdd{biswap@visva-bharati.ac.in}

\abstract{We propose a method for testing homogeneity in three
  dimensional spatial distributions using Renyi entropy. We apply the
  proposed method to data from cosmological N-body simulations and
  Monte Carlo simulations of homogeneous Poisson point process. We
  show that the method can effectively characterize the
  inhomogeneities and identify any transition scale to homogeneity, if
  present in such distributions. The proposed method can be used to
  study the cosmic homogeneity in present and future generation galaxy
  redshift surveys.}

\begin{document}
\maketitle
\flushbottom

\section{Introduction}
The assumption of statistical homogeneity and isotropy of the Universe
on sufficiently large scales, known as the Cosmological principle, is
fundamental to modern cosmology. The principle can not be proved in a
strictly mathematical sense and was introduced in cosmology largely
due to its great aesthetic appeal and simplicity. It allows us to
mathematically describe the global structure of the Universe using FRW
space-time geometry. We rely on the FRW geometry while analyzing and
interpreting data from various cosmological observations. So the
assumption is of paramount importance for our current understanding of
the Universe and the validity of this assumption must be tested with
different observations. Besides, inhomogeneities may also play an
important role in explaining the observed cosmic acceleration through
the backreaction mechanism \citep{buchert97, schwarz, kolb06,
  buchert08,ellis}.

The isotropy of the Universe is supported by a multitude of
observations such as CMBR \citep{penzias,smoot,fixsen}, radio sources
\citep{wilson,blake}, X-ray background \citep{peeb93,wu,scharf}, Gamma
ray bursts \citep{meegan,briggs}, supernovae \citep{gupta,lin} and
galaxies \citep{marinoni,alonso,sarkariso19}. However these
observations alone can not assert the large-scale statistical
homogeneity of the Universe. Such a validation is only possible if we
believe that our location in the Universe is not a special one.

The observed galaxy distribution is known to exhibit scale invariant
features on small scales \citep{pietronero,coleman92,mandelbrot} which
resembles fractals. A number of studies
\citep{pietronero,coleman92,amen,joyce,labini07, labini09, labini11}
claim that such scale-invariant behaviour continues on larger length
scales extending out to the scale of the surveys, which indicates that
there are no transition scale to homogeneity. Many other studies
reaffirm the scale invariant nature of galaxy distribution on small
scales but most of them
\citep{martinez94,borgani95,guzzo97,cappi,bharad99,pan2000,yadav,hogg,prakash,scrim,nadathur,pandeysarkar15,pandeysarkar16}
reported a transition to homogeneity on scales $70-150 \hmpc$.

Various observations point out to the existence of structures in the
Universe, which extend up to several hundreds of Mpc. The Sloan Great
Wall (SGW) in the SDSS galaxy distribution is known to extend over
length scales of $\sim 400$ Mpc \citep{gott05}. The large quasar
groups (LQG) in the quasar distribution at $z\sim 1.3$ is known to
have a characteristic size of $\sim 500 \hmpc$ \citep{clowes}. The
Eridanus supervoid is believed to stretch across a region, which
extends up to $\sim 300$ Mpc \citep{szapudi}. The existence of such
large-scale structures may challenge the validity of the cosmological
principle and the standard cosmological model. Using Horizon Run 2
simulation, Park et al. \citep{park12} show that existence of high
density and low density regions of such extent in observations are
consistent with $\Lambda$CDM paradigm. It is also important to address
the statistical significance of any such structures identified in
observations.  For instance, Nadathur \cite{nadathur} pointed out that
the algorithm used for identification of LQGs yield even larger
structures in simulations of a homogeneous Poisson point process.

Most of the traditional methods for testing homogeneity are based on
the number counts in spheres centered around galaxies. Multi-fractal
analysis \citep{martinez90, coleman92, borgani95, bharad99, yadav} of
galaxies is one of the most widely used method for testing cosmic
homogeneity. It characterizes The scale of homogeneity by studying the
scaling of different moments of number counts. In a multi-fractal,
different moments of the distribution scale with different scaling
exponent. The multi-fractals can be defined based on the Renyi
dimension or generalized dimension \citep{renyi70, hentschel}. However
$r \to 0$ limit in these definitions are not meaningful for observed
galaxy distributions and it is difficult to measure them accurately
\citep{saslaw99}. Pandey \cite{pandey13} defined a statistical measure
for homogeneity based on the Shannon entropy \citep{shannon48} and
used it to measure the scale of homogeneity in the Main Galaxy sample
\citep{pandeysarkar15}, LRG sample \citep{pandeysarkar16} and BOSS
sample \citep{sarkarpandey16} from the SDSS \citep{york}. In
information theory, Renyi entropy \citep{renyi61} is one of the
families of functionals which quantify the uncertainty or randomness
of a system. The Shannon entropy is the limiting case of the Renyi
entropy.  The Renyi entropies of higher order are more sensitive to
the presence of inhomogeneities in a distribution. In the present
work, we propose a more general statistical measure of homogeneity
based on the Renyi entropy. We apply the proposed method to data from
simulations of homogeneous Poisson point process and distributions of
particles from N-body simulations. We explore the scope and
limitations of the proposed method in studying cosmic homogeneity
using the present generation and forthcoming galaxy surveys.

The outline of the paper is as follows: We explain the method of
analysis in Section 2 and describe the data in Section 3. We
discuss the results and present our conclusions in Section 4.

\section{Method of Analysis}
Information theory is an interdisciplinary branch of science which
owes its origin to a seminal paper \citep{shannon48} by Claude Shannon.

Shannon entropy measures the average information content of a random
variable. The Shannon entropy of a discrete random variable $X$ is
defined as,\\
\begin{eqnarray}
H(X) & = & - \sum^{n}_{i=1} \, p(x_{i}) \, \log \, p(x_{i})
\label{eq:one}
\end{eqnarray}
,where $p(x_i)$ is the probability of $i^{th}$ event out of a
total $n$ outcomes $\{x_{i}:i=1,....n\}$.

The Renyi entropy \citep{renyi61} generalizes Shannon entropy, which was
originally proposed by Alfred Renyi in 1961. The Renyi entropy of
order $q$ for a random variable $X$ is defined as,\\
\begin{eqnarray}
S_q(X) & = & \frac{1}{1-q}\, \log \sum^{n}_{i=1} \, p^q(x_{i})
\label{eq:two}
\end{eqnarray}
,where $q \in [0,\infty]$. For $q=0$, we get the maximum entropy which
is the logarithm of the size of the support of $p$. The expression
given in equation 2.2 is potentially undefined for $q=1$. Applying
L'Hospital's rule, one can show that the expression for Renyi entropy
in equation 2.2 reduces to Shannon entropy for $q=1$.

$S_q$ is a weakly decreasing function of $q$. Renyi entropy weights
the probabilities in a non-uniform manner. Regardless of their values,
the probabilities are weighted more equally for lower values of
$q$. On the other hand, Renyi entropy for higher values of $q$ are
increasingly determined by the higher probability events. If all the
probabilities are equal then Renyi entropies have the same value
$S_q(X)=\log n$ irrespective of their order.

The Renyi dimension or the generalized dimension $D_q$ \citep{renyi70}
of order $q$ is defined as,\\
\begin{eqnarray}
D_q(X) & = & \lim_{\epsilon \to 0} \frac{S_q(X)}{\log \frac{1}{\epsilon}} 
\label{eq:three}
\end{eqnarray}
, where $\epsilon$ is the scaling factor.

One can define Shannon information dimension in an analogous manner by
replacing Renyi entropy with Shannon entropy $H(X)$ in equation
2.3. It may be noted that Shannon information dimension reduces to
fractal dimension for a discrete uniform distribution where $H(X)=\log
n$.

Renyi dimension or generalized dimension are often used to test
homogeneity in galaxy distributions. One particular disadvantage of
the measure is that $\epsilon \to 0$ limit in these definitions are
not meaningful for the observed galaxy distributions.  So the spectrum
of generalized dimensions obtained in this limit are not
accurate. Further, these measures are usually evaluated using a finite
number of galaxies. In reality, a stable and correct estimate of the
spectrum requires a much larger number of galaxies than those are
generally available in a volume limited sample.

In the present work, we propose a simple measure of homogeneity for
galaxy distribution based on the fact that Renyi entropies of
different order assumes the same value when the probabilities of all
the outcomes are equal. Let us assume $N$ galaxies distributed within
a volume $V$. We consider each of the galaxies and consider a sphere
of radius $r$ centered on it.  The number of galaxies $n_i(<r)$ within
the sphere around the $i^{th}$ galaxy is given by,\\
\begin{eqnarray}
 n_i(<r)=\sum_{j=1}^{N}\Theta(r-\mid {\bf{x}}_i-\bf{x}_j \mid)
\label{eq:four}
\end{eqnarray}
, where ${\bf{x}_{i}}$ and ${\bf{x}_{j}}$ in the Heaviside step
function $\Theta$ are the radius vector of $i^{th}$ and $j^{th}$
galaxies respectively. We take into account the edge effects by
discarding all the galaxies which lie closer than $r$ from the
boundary of the volume. We define a random variable $X_r$
corresponding to radius $r$. Only a finite number of valid centres
$M(r)$ would be available at a radius $r$ and the number of such valid
centre would decrease with increasing radius due to the finite volume
of the distribution. The probability of finding another galaxy within
a distance $r$ from a galaxy is directly proportional to the number of
galaxies within a sphere of radius $r$ around it. Now let us consider
the subset of valid centres at a given radius. If a galaxy is randomly
picked up from the $M(r)$ centres available at radius $r$, we have
$M(r)$ possible outcomes for this event. The probability of randomly
selecting the $i^{th}$ centre is given by,
$f_{i,r}=\frac{\rho_{i,r}}{\sum^{M(r)}_{i=1} \, \rho_{i,r}}$ , where
the density at the location of $i^{th}$ center is
$\rho_{i,r}=\frac{n_{i}(<r)}{\frac{4}{3}\pi r^{3}}$. We can write
$\sum^{M(r)}_{i=1} \, f_{i,r}=1$ which implies that the sum of the
probabilities from all the outcomes of the event is 1.

The Renyi entropy of order $q$ associated with the random variable
$X_{r}$ can be written as,\\
\begin{eqnarray}
S_{q}(r) & = & \frac{1}{1-q} \log \sum_{i=1}^{M(r)} \, f^q_{i,r} \nonumber\\
& = & \frac{1}{1-q} \log \frac{\sum_{i=1}^{M(r)} n_i^q(<r)}{(\sum^{M(r)}_{i=1} n_i(<r))^q} 
\label{eq:five}
\end{eqnarray}\\
We choose the base of the logarithm in the above formula to be $10$. 

In an ideal homogeneous distribution, all the spheres around the
$M(r)$ centres would contain exactly same number of galaxies within
them. Such a distribution would maximize the uncertainty in the random
variable $X_r$ as the probabilities of selection for each and every
centre would be same. When $f_{i,r}=\frac{1}{M(r)}$ for all the
centres then all the Renyi entropies of different orders reduce to
$S_q(r)=\log \,M(r)$ which we label as $[S_q(r)]_{max}$. We calculate
the ratio $\frac{S_q(r)}{[S_q(r)]_{max}}$ to normalize the Renyi
entropies of different order by the maximum possible entropy at any
given length scale. The Renyi entropy has a remarkable advantage over
Shannon entropy as a measure of homogeneity. Since the Renyi entropies
of higher order assign progressively greater weights to higher
probabilities, they would be more sensitive to the presence of
inhomogeneities in the distribution. In general, one can consider
Renyi entropies up to any order. We use the Renyi entropies up to
order of $10$ keeping in mind the finite and discrete nature of the
distributions. The Renyi entropies of different orders may not be
exactly equal as we are working with finite and discrete
distributions. We define the scale of homogeneity as the scale at
which all the Renyi entropies of different order are nearly equal and
the quantity $1-\frac{S_q(r)}{[S_q(r)]_{max}}$ for all $q$ are smaller
than $10^{-3}$.

\section{Data}
We apply the proposed method to data from N-body simulations and
simulations of homogeneous Poisson point process. We describe the
preparation of data in the following subsections.

\subsection{N-body simulations}

We analyze data from a set of N-body simulations of the present day
($z=0$) distributions of dark matter. These simulations were run for a
previous study by Pandey \citep{pandey13}. A Particle-Mesh (PM) N-body
code was used to simulate the distributions of $256^3$ particles on a
$512^3$ mesh which occupy a comoving volume of $[921.6 \hmpc]^3$. The
simulations used $\Omega_{m0}=0.27, \Omega_{\Lambda0}=0.73 ,h=0.71$
and a $\Lambda$CDM power spectrum with $n_s=0.96$ and $\sigma_8=0.812$
\citep{komatsu}. We run these simulations for 3 different realizations
of the initial density perturbations. We also prepare a set of
distributions which are biased relative to the dark matter
distributions. We employ the ``sharp cutoff'' biasing scheme
\citep{cole} to generate the biased distributions from the original
dark matter distributions. Particles are selected by applying a sharp
cut-off to the smoothed density field of dark matter
distributions. One can label these selected particles as galaxies. The
linear bias parameter $b$ of the simulated biased distributions is
determined by,\\
\begin{eqnarray}
b=\sqrt{ \frac {\xi_g(r)}{\xi_{dm}(r)}}
\label{eq:six}
\end{eqnarray}
, where $\xi_g(r)$ and $\xi_{dm}(r)$ are respectively the two-point
correlation functions of galaxy and dark matter distributions. We
simulate a set of biased distributions with linear bias parameter
$b=2$. The original dark matter distributions are unbiased and have a
linear bias of $b=1$.

For the present analysis, we identify three non overlapping spherical
regions of radius $R=200 \hmpc$ from each of the simulations and
randomly extract $N=10^5$ particles within each of them. This gives us
a total 9 such samples for each bias values.

\subsection{Simulations of homogeneous Poisson point process}
 
 We generate a set of Monte Carlo realizations of a homogeneous
 Poisson point process. The homogeneous Poisson point process will
 have constant density everywhere. A radial density function can be
 mapped to a probability function by normalizing it to one within
 interval $r=0$ to $r=R$. The desired number of particles are enforced
 to be distributed within radius $R$. The probability of finding a
 particle at a given radius $r$ is proportional to the density at that
 radius, which can be expressed as, $P(r)=\frac{ r^2 \lambda(r) }
 {\int_{0}^{R} r^2 \lambda(r) \, dr}$. Here $\lambda(r)$ describes the
 radial variation in density which can have different functional form
 for inhomogeneous Poisson point processes. For a homogeneous Poisson
 point process $\lambda(r) = 1$ as the density is independent of
 location. The maxima of $P(r)$ in this case is at $r=R$, which we
 label as $P_{max}$. We randomly choose a radius within $0 \le r \le
 R$ and calculate the probability of finding a particle at that radius
 using the expression for $P(r)$. We then randomly generate a
 probability value $P(r)$ in the range $0 \le P(r) \le P_{max}$. We
 accept the radius $r$ only if the calculated value of $P(r)$ is
 greater than its randomly selected value. The selected radius is then
 assigned isotropically selected angular co-ordinates $\theta$ and
 $\phi$. We choose the radius of the spherical region to be $R=200
 \hmpc$ and number of points within to be $N=10^{5}$. We simulate $10$
 such realizations for the present study.

\begin{figure*}[htbp!]
\centering
\includegraphics[width=12cm]{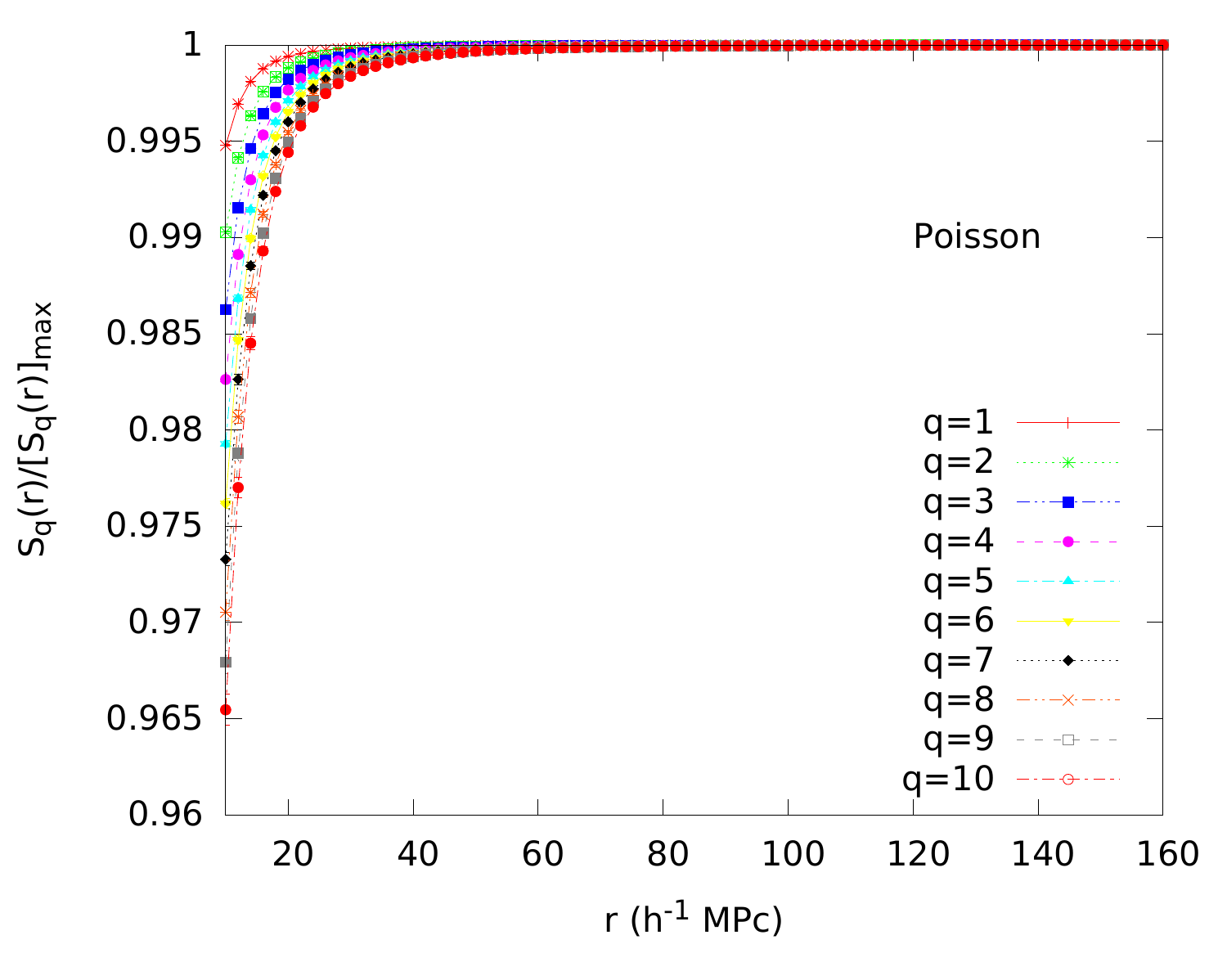}
\caption{This shows $\frac{S_q(r)}{[S_q(r)]_{max}}$ as a function of
  length scale $r$ for a homogeneous Poisson point process. The
  results for different $q$ values are shown together for a
  comparison. The $1-\sigma$ errors bars at each data points are
  obtained from 10 different realizations.}
\label{poisson}
\end{figure*}

\begin{figure*}[htbp!]
\centering
\includegraphics[width=12cm]{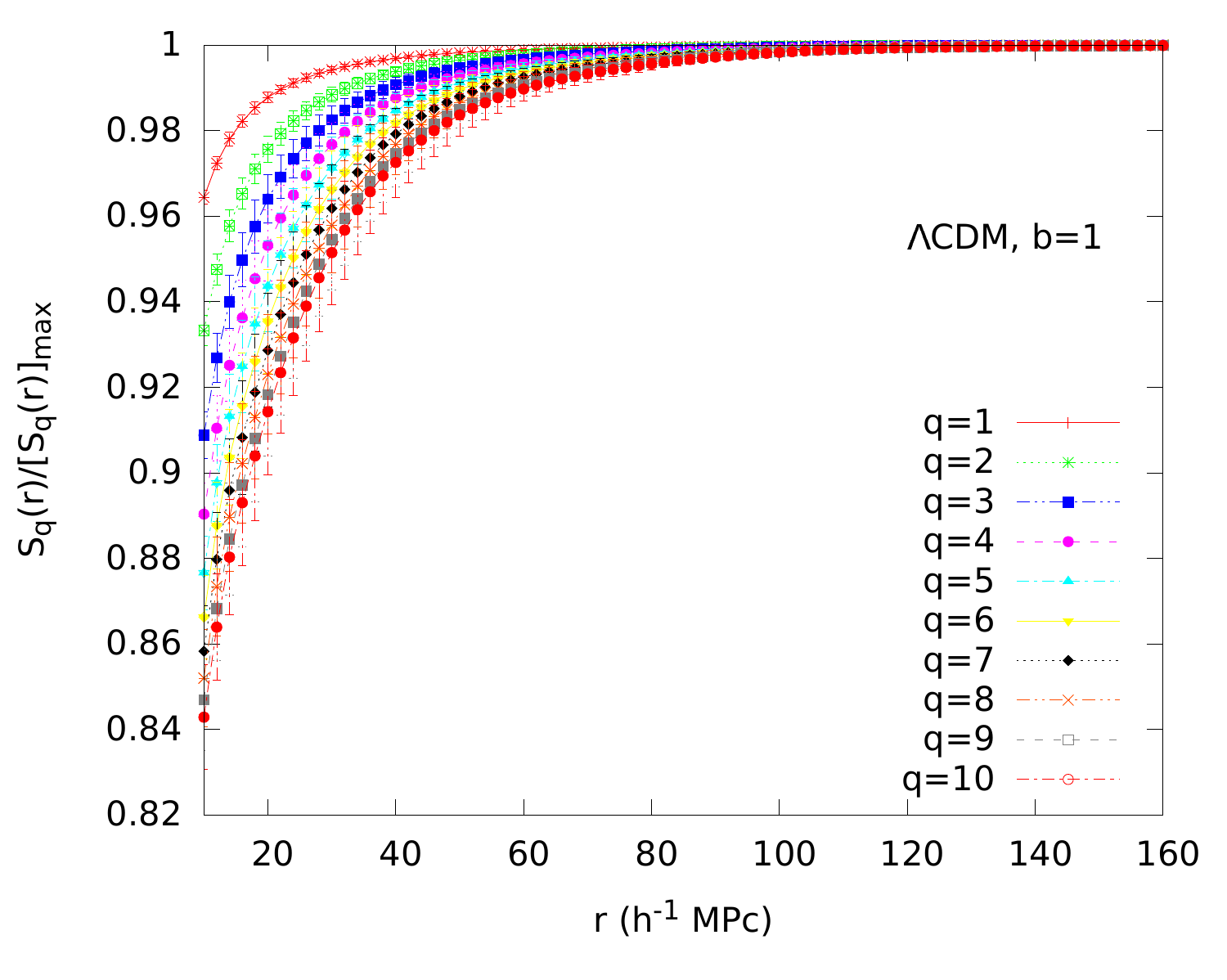}
\caption{Same as Figure 1 but for unbiased $\Lambda$CDM model.}
\label{nbodyb1}
\end{figure*}

\begin{figure*}[htbp!]
\centering
\includegraphics[width=12cm]{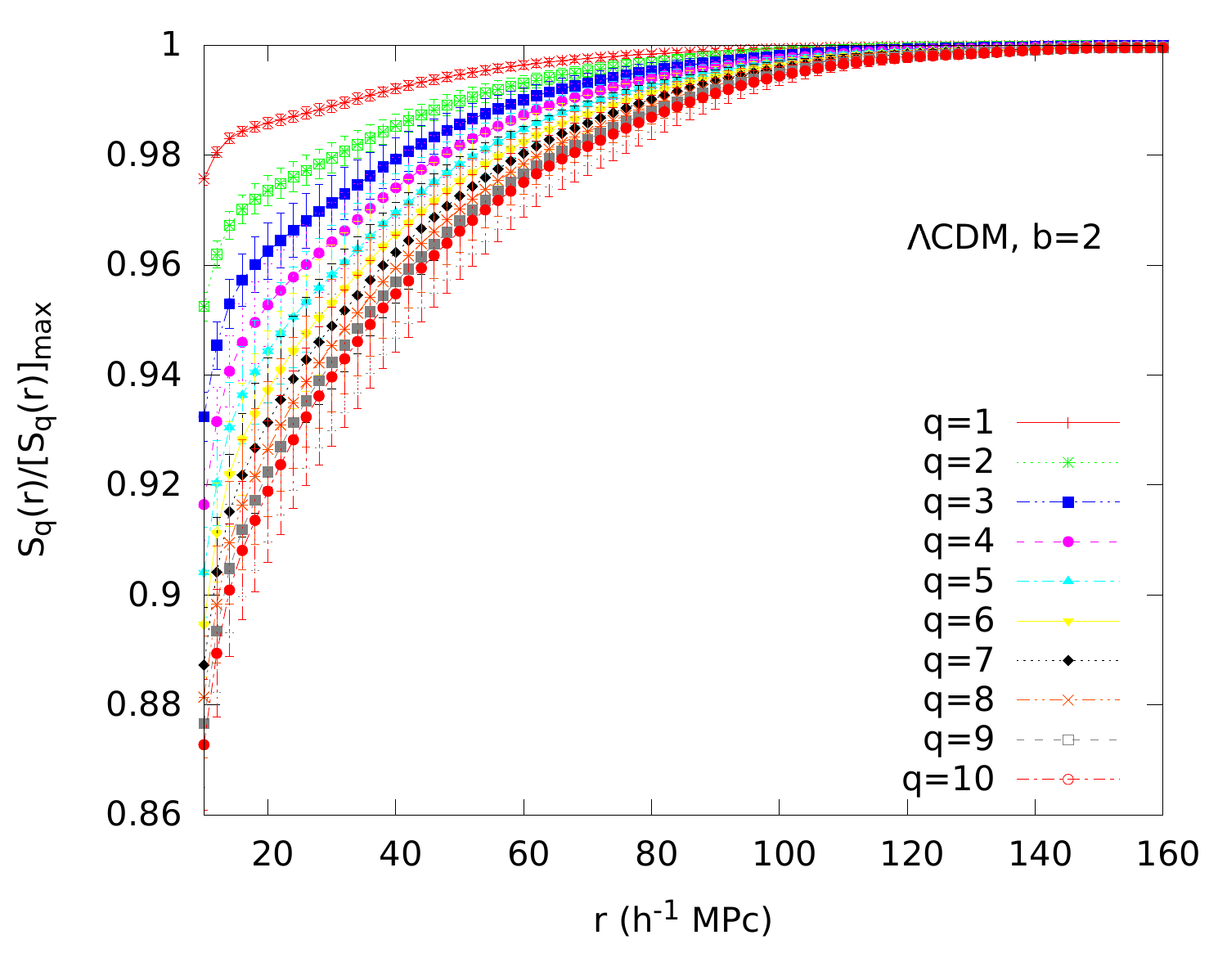}
\caption{Same as Figure 1 but for $\Lambda$CDM model with $b=2$.}
\label{nbodyb2}
\end{figure*}

\section{Results and Conclusions}

We show the Renyi entropies of different order for a homogeneous
Poisson point process in Figure \ref{poisson}. The ratio
$\frac{S_q(r)}{[S_q(r)]_{max}}$ for $10$ different $q$ values are
shown together as a function of length scale $r$. We see that the
ratio deviates from $1$ for all the $q$ values at the smallest length
scale $r=10 \hmpc$, the deviation being largest for $q=10$. The larger
deviation for higher $q$ values are related to the fact that Renyi
entropy is a slowly decreasing function of $q$. The deviations
indicate that the distribution is inhomogeneous on $10 \hmpc$. The
deviations gradually diminish with increasing radius $r$ for all the
$q$ values and the quantity $1-\frac{S_q(r)}{[S_q(r)]_{max}}$ drops
below $10^{-3}$ for all of them on a length scale of $50 \hmpc$. These
inhomogeneities are an outcome of the discrete nature of the
distributions. All the Poisson distributions analyzed here show a
transition to homogeneity on $50 \hmpc$. The $1-\sigma$ errorbars
shown at each data point are drawn from 10 different realizations.

We show the results for the unbiased $\Lambda$CDM model in Figure
\ref{nbodyb1}.  The Figure \ref{nbodyb1} show a similar trend as
observed in Figure \ref{poisson}. However, the degree of
inhomogeneities and their variations with length scale in the
$\Lambda$CDM model are noticeably different than observed in the
Poisson distributions. In this figure, the ratio
$\frac{S_q(r)}{[S_q(r)]_{max}}$ for each values of $q$ shows a larger
deviation from 1, as compared to their values observed in Poisson
distributions. Further, the inhomogeneities in the $\Lambda$CDM model
decrease much slowly with radius $r$ as compared to Poisson
distributions. The higher degree of inhomogeneities and their slower
variations in the $\Lambda$CDM model arise due to the existence of
real inhomogeneities in the distribution. We find that
$\frac{S_q(r)}{[S_q(r)]_{max}}$ for all the $q$ values converges to
1 within $10^{-3}$ at a length scale of $120 \hmpc$. The results
indicate a transition to homogeneity at $120 \hmpc$ for the
$\Lambda$CDM model. The $1-\sigma$ errorbars at each data point are
obtained from 9 different realizations.

We repeat our analysis with the data from the $\Lambda$CDM model with
$b=2$ and show the corresponding results in Figure \ref{nbodyb2}. We
find that the observed inhomogeneities in the biased $\Lambda$CDM
model also diminish with increasing length scales. Interestingly, the
degree of inhomogeneities in the biased $\Lambda$CDM model are lower
than the unbiased $\Lambda$CDM model for all $q$ values at smaller
length scales. However, the inhomogeneities in the biased $\Lambda$CDM
model are larger than its unbiased counterpart on larger length
scales. The matter distribution is known to exhibit a weblike network
of nodes, filaments and sheets surrounded by voids. A biased
distribution is primarily composed of particles preferentially
identified from rarer density peaks which represent similar
environments. So there will be less disparity in the number counts
around the centres located in such environments. This is particularly
true on smaller length scales up to the physical extent of these
regions. But the measurements around these centres would show a larger
disparity beyond the extent of these environments On the other hand,
the particles are distributed across diverse environments in the
unbiased $\Lambda$CDM model. The measuring spheres centred on the
particles in such a distribution trace diverse environments giving
rise to a larger disparity in their measurements. The disparity in
these measurements gradually decrease with increasing length scales
until the measuring spheres include statistically similar number of
nodes, filaments, sheets and voids. Figure \ref{nbodyb2} shows that
the biased $\Lambda$CDM model show a slower variation of
inhomogeneities with length scales than the unbiased
model. Consequently, the inhomogeneities extend to a larger length
scales in the biased model. We find that the quantity
$\frac{S_q(r)}{[S_q(r)]_{max}}$ for all the $q$ values in the
$\Lambda$CDM model with $b=2$ converges to 1 within $10^{-3}$ at a
length scale of $140 \hmpc$. We obtain the $1-\sigma$ errorbars at
each data point using 10 different realizations.

One particular disadvantage of any number count based method is that
the measuring centres progressively get confined towards the centre of
the survey volume with increasing length scales. This confinement bias
\citep{pandey13, kraljic} enforces overlaps between the measuring
spheres leading towards an apparent homogeneity in any inhomogeneous
distributions. The effects of confinement bias can be minimized by
increasing the survey volume. However larger survey volume requires us
to take in to account the evolution with look back time. This could
make anti-Copernican void models to appear homogeneous on larger
length scales. Fortunately, other observations like SNe, CMB and BAO
can be used to constrain such models \citep{zibin, clifton, biswas,
  chris}.

The method proposed in this work has some similarities to the other
studies on homogeneity that are based on multifractal analysis
\citep{yadav, prakash, scrim}. Both of these methods are based on the
counts-in-spheres statistics. However, there are some important
differences between the two approach. In the multifractal analysis,
the Minkowski-Bouligand dimensions $D_q$ are estimated from the
logarithmic slope of the generalised correlation integral. The
generalized dimension $D_q$ is independent of $q$ for a mono fractal
but depends on the values of $q$ in a multifractal. A monofractal is
regarded as homogeneous when $D_q$ becomes equal to the ambient
dimension irrespective of the order $q$. The positive values of $q$
assign greater weights to the overdense regions whereas negative
values put more weights to the underdense regions. The generalized
dimensions $D_q$ for $q<0$ are known to be very sensitive to the low
density regions in finite datasets \citep{roberts}. This makes it
difficult to distinguish empty space from the space filled with matter
at low probability, which may dramatically affect $D_q$ for the
negative $q$ values. The Renyi entropy based method proposed in this
work uses only positive values of $q$ where smaller $q$ values weights
the probabilities in a more uniform manner and higher $q$ values puts
more weights to higher probabilities. The measurements of the spectrum
of generalized dimension requires us to calculate the derivative of
the generalized correlation integral whereas the present method does
not involve any numerical derivatives and hence bypasses the
associated errors. The multifractal analysis determines the
homogeneity scale based on the scaling of the different moments of
galaxy counts, whereas the present method identifies the homogeneity
scale based on the maximization of uncertainty in the random variable
measuring the density of points. The Renyi entropy is a generalization
of the Shannon entropy, which quantify the uncertainty or randomness
of a system. The proposed method is based on the simple fact that all
the Renyi entropies of different orders must have the same value when
the probabilities are equal. We use the normalized Renyi entropies for
our analysis which would be less susceptible to the survey geometry
and incompleteness effects \citep{pandeysarkar15}. On the other hand,
the generalized dimension $D_q$ is known to be more susceptive to
these issues \citep{scrim, avila}.

The cosmological N-body simulations are performed with periodic
boundary conditions to mimic the large-scale homogeneity of the matter
distribution in the Universe. The periodicity in simulations are
typically assumed on the scale of the box. In the present analysis,
the simulations were carried out within a comoving volume of $[921.6
  \, h^{-1}\, {\rm Mpc}]^3$ and the scale of homogeneity measured in
these simulations lie in the range $120-140 \, h^{-1}\, {\rm
  Mpc}$. These are much smaller than the size of the simulation
boxes. We note that the homogeneity scale measured in the homogeneous
Poisson distributions are $\sim 50 \, h^{-1}\, {\rm Mpc}$ which is
significantly smaller than those measured in the unbiased and biased
simulations of the $\Lambda$CDM model. Also, the degree of
inhomogeneity in the Poisson distributions are reasonably smaller than
the distributions derived from the N-body simulations. The Poisson
distribution is homogeneous by construction and is expected to exhibit
homogeneity on a smaller scale. The small inhomogeneities measured in
the Poisson distributions are an outcome of the discreteness noise
that diminishes with increasing length scales. Since the measured
homogeneity scale in the simulations are significantly smaller than
the assumed homogeneity scale, we can accept the measured homogeneity
scales as reliable.

We show that the statistical measure proposed in this work can
effectively quantify the inhomogeneities present in different types of
distributions. The measure can also characterize the nature of
inhomogeneities and detect the transition scale to homogeneity, if
present in a given distribution. The proposed measure can be used to
test the assumption of cosmic homogeneity in the present and future
generation galaxy surveys.

\section*{Acknowledgments}
I acknowledge financial support from the SERB, DST, Government of
India through the project CRG/2019/001110. I would also like to
acknowledge IUCAA, Pune for providing support through associateship
programme. The author acknowledges the Computing Center of the Max
Planck Society in Garching (RZG) for the computing facilities provided
for the simulations used in this work.


\begin{thebibliography}{99}
\bibitem{buchert97} T. Buchert, \& J. Ehlers, \aap, \textbf{320}, 1
(1997)
\bibitem{schwarz} D.~J. Schwarz, arXiv:astro-ph/0209584 (2002)
\bibitem{kolb06} E.~W. Kolb, S. Matarrese \& A. Riotto, New Journal of
Physics, \textbf{8}, 322 (2006)
\bibitem{buchert08} T. Buchert, General Relativity and Gravitation,
\textbf{40}, 467 (2008)
\bibitem{ellis} G.~F.~R. Ellis, Classical and Quantum Gravity,
\textbf{28}, 164001 (2011)
\bibitem{penzias} A.~A. Penzias \& R.~W. Wilson, \apj, \textbf{142},
419 (1965)
\bibitem{smoot} G.~F. Smoot, C.~L. Bennett, A. Kogut, et al., \apjl,
\textbf{396}, L1 (1992)
\bibitem{fixsen} D.~J. Fixsen, E.~S. Cheng, J.~M. Gales, et al., \apj,
\textbf{473}, 576 (1996)
\bibitem{wilson} R.~W. Wilson \& A.~A. Penzias, Science, \textbf{156},
1100 (1967)
\bibitem{blake} C. Blake \& J. Wall, \nat, \textbf{416}, 150 (2002)
\bibitem{peeb93} P.~J.~E. Peebles, Principles of Physical
Cosmology.~Princeton, N.J., Princeton University Press (1993)
\bibitem{wu} K.~K.~S. Wu, O. Lahav \& M.~J. Rees, \nat, \textbf{397},
225 (1999)
\bibitem{scharf} C.~A. Scharf, K. Jahoda, M. Treyer, et al., \apj,
\textbf{544}, 49 (2000)
\bibitem{meegan} C.~A. Meegan, G.~J. Fishman, , R.~B. Wilson, et al.,
\nat, \textbf{355}, 143 (1992)
\bibitem{briggs} M.~S. Briggs, W.~S. Paciesas, G.~N.  Pendleton, et
al., \apj, \textbf{459}, 40 (1996)
\bibitem{gupta} S. Gupta \& T.~D. Saini, \mnras, \textbf{407}, 651
(2010)
\bibitem{lin} H.-N. Lin, S. Wang, Z. Chang \& X.Li,\mnras,
\textbf{456}, 1881 (2016)
\bibitem{marinoni} C. Marinoni, J. Bel \& A. Buzzi, \jcap,
\textbf{10}, 036 (2012)
\bibitem{alonso} D. Alonso, A.~I. Salvador, F.~J.  S{\'a}nchez, et
al., \mnras, \textbf{449}, 670 (2015)
\bibitem{sarkariso19} S. Sarkar, B. Pandey, R. Khatri, MNRAS,
\textbf{483}, 2453 (2019)
\bibitem{pietronero} L. Pietronero, Physica A Statistical Mechanics
  and its Applications, \textbf{144}, 257 (1987)
\bibitem{coleman92} P.~H. Coleman, L. Pietronero, \physrep,
\textbf{213}, 311 (1992)
\bibitem{mandelbrot} B.~B. Mandelbrot, Astrophysical Letters and
  Communications, \textbf{36}, 1 (1997)
\bibitem{amen} L. Amendola,\& E. Palladino, \apjl, \textbf{514}, L1
  (1999)
\bibitem{joyce} M. Joyce, M. Montuori \& F.~S. Labini, \apjl,
  \textbf{514}, L5 (1999)
\bibitem{labini07} F. Sylos Labini, N.~L. Vasilyev \& Y.~V. Baryshev,
  \aap, \textbf{465}, 23 (2007)
\bibitem{labini09} F. Sylos Labini, N.~L. Vasilyev \& Y.~V. Baryshev,
  \aap, \textbf{508}, 17 (2009)
\bibitem{labini11} F. Sylos Labini, Europhysics Letters, \textbf{96},
  59001 (2011)
\bibitem{martinez94} V.~J. Martinez \& P. Coles, \apj, \textbf{437},
550 (1994)
\bibitem{borgani95} S. Borgani, \physrep, \textbf{251}, 1 (1995)
\bibitem{guzzo97}  L. Guzzo, New Astronomy, \textbf{2}, 517 (1997)
\bibitem{cappi} A. Cappi, C. Benoist, L.~N. da Costa, \&
  S. Maurogordato, \aap, \textbf{335}, 779 (1998)
\bibitem{bharad99} S. Bharadwaj, A.~K. Gupta \& T.~R. Seshadri, \aap,
\textbf{351}, 405 (1999)
\bibitem{pan2000} J. Pan \& P. Coles, \mnras, \textbf{318}, L51 (2000)
\bibitem{yadav} J. Yadav, S. Bharadwaj, B. Pandey \& T.~R. Seshadri,
\mnras,\textbf{364}, 601 (2005)
\bibitem{hogg} D.~W. Hogg, D.~J. Eisenstein, M.~R. Blanton,
  N.~A. Bahcall, J. Brinkmann, J.~E. Gunn \& D.~P. Schneider, \apj,
  \textbf{624}, 54 (2005)
\bibitem{prakash} P. Sarkar, J. Yadav, B. Pandey \& S. Bharadwaj,
  \mnras, \textbf{399}, L128 (2009)
\bibitem{scrim}  M.~I. Scrimgeour,  T. Davis, C. Blake, et al.\ 2012,
  \mnras, 3412 (2012)
\bibitem{nadathur} S. Nadathur, \mnras, \textbf{434}, 398 (2013)
\bibitem{pandeysarkar15} B. Pandey, S. Sarkar, MNRAS, \textbf{454},
2647 (2015)
\bibitem{pandeysarkar16} B. Pandey, S. Sarkar, MNRAS, \textbf{460},
1519 (2016)
\bibitem{gott05} J.~R., III, Gott, M. Juri{\'c}, D. Schlegel, et al.,
\apj, \textbf{624}, 463 (2005)
\bibitem{clowes} R.~G. Clowes, K.~A. Harris, S. Raghunathan, et al.,
\mnras, 429, 2910 (2013)
\bibitem{szapudi} I. Szapudi, A. Kov{\'a}cs, B.~R. Granett, et al.,
\mnras, \textbf{450}, 288 (2015)
\bibitem{park12} C. Park, Y.-Y. Choi, J. Kim, J.~R. Gott , S.~S. Kim ,
K.-S. Kim, ApJL, \textbf{759}, L7 (2012)
\bibitem{martinez90} V.~J. Martinez \& B.~J.~T. Jones, \mnras,
\textbf{242}, 517 (1990)
\bibitem{renyi70} A. Renyi, Probability Theory, Published by
North-Holland Publishing Company, Amsterdam (1970)
\bibitem{hentschel} H.G.E. Hentschel, I. Procaccia, Physica
\textbf{8D}, 435 (1983)
\bibitem{saslaw99} W.~C. Saslaw, The Distribution of the
Galaxies. Published by Cambridge University Press, (1999)
\bibitem{pandey13} B. Pandey, MNRAS, \textbf{430}, 3376 (2013)
\bibitem{shannon48} C. E. Shannon, Bell System Technical Journal,
\textbf{27}, 379 (1948)
\bibitem{sarkarpandey16} S. Sarkar, B. Pandey, MNRAS, \textbf{463},
L12 (2016)
\bibitem{york} D.~G. York, et al., \aj, \textbf{120}, 1579 (2000)
\bibitem{renyi61} A. Renyi, Proceedings of the fourth Berkeley
Symposium on Mathematics, Statistics and Probability, pp. 547-561
(1961)
\bibitem{komatsu} E. Komatsu, et al., \apjs, \textbf{180}, 330 (2009)
\bibitem{cole} S. Cole, S. Hatton, D.~H. Weinberg, , \& C.~S. Frenk,
  \mnras, \textbf{300}, 945 (1998)
\bibitem{kraljic} D. Kraljic, \mnras, \textbf{451}, 3393 (2015)
\bibitem{zibin} J.~P. Zibin, A. Moss \& D. Scott, Physical Review
  Letters, \textbf{101}, 251303 (2008)
\bibitem{clifton} T. Clifton, P.~G.Ferreira, , \& K. Land, Physical
  Review Letters, \textbf{101}, 131302 (2008)
\bibitem{biswas} T. Biswas, A. Notari, \&  W. Valkenburg, \jcap, \textbf{11}, 30 (2010)
\bibitem{chris} C. Clarkson, Comptes Rendus Physique, \textbf{13}, 682
  (2012)
\bibitem{avila} F. Avila, C.~P. Novaes, A. Bernui, E. de Carvalho, J.~P. Nogueira-Cavalcante, \mnras, \textbf{488}, 1481 (2019) 
\bibitem{roberts} A.~J. Roberts , 2005, arXiv, nlin/0512014
\end{thebibliography}
\end{document}